\documentclass[11pt ]{article}
\usepackage{amsmath}
\usepackage{graphicx}
\usepackage{hyperref}
\usepackage{lmodern}
\usepackage{amssymb}
\usepackage{booktabs}
\usepackage{tikz}
\usetikzlibrary{patterns}
\renewcommand{\theequation}
{\arabic{section}.\arabic{equation}}

\makeatletter
\def\eqnarray{ \stepcounter{equation} \let\@currentlabel=\theequation
 \global\@eqnswtrue
 \global\@eqcnt\z@
 \tabskip\@centering
 \let\\=\@eqncr
 $$\halign to \displaywidth\bgroup\@eqnsel\hskip\@centering
 $\displaystyle\tabskip\z@{##}$&\global\@eqcnt\@ne
 \hfil$\displaystyle{{}##{}}$\hfil
 &\global\@eqcnt\tw@$\displaystyle\tabskip\z@{##}$\hfil
 \tabskip\@centering&\llap{##}\tabskip\z@\cr}
\makeatother

\makeatletter
\def\@arrayacol{\edef\@preamble{\@preamble \hskip .5\arraycolsep}}
\def\array{\let\@acol\@arrayacol \let\@classz\@arrayclassz
\let\@classiv\@arrayclassiv \let\\\@arraycr\def\@halignto{}\@tabarray}
\makeatother

\makeatletter
\newcounter{subeqncnt}
\def\thesubeqncnt{\alph{subeqncnt}}
\def\subequations{\begingroup%
   \stepcounter{equation}\edef\@tempa{\theequation}%
   \let\c@equation\c@subeqncnt\c@subeqncnt\z@
   \edef\theequation{\@tempa\noexpand\thesubeqncnt}}

\makeatother

\newcommand{\be}{\begin{equation}}
\newcommand{\ee}{\end{equation}}

\newcommand{\beqa}{\begin{eqnarray}}
\newcommand{\eeqa}{\end{eqnarray}}

\begin{document}


\centerline{STRING FINE TUNING} \vskip 4cm

\centerline{G.Savvidy,~K.Savvidy} \bigskip

\centerline{Institut f\"ur Theoretische Physik der Universit\"at
Frankfurt,}
\centerline{D-6000 Frankfurt am Main 11,Fed.Rep.Germany}
\vskip 3cm

{\obeylines\everypar{\hfill}
..hat man die eine,
die synthetische (Methode),
sehr vernachl\"assigt...
Jacob Steiner.} \vskip 3cm

\centerline{ Abstract}\medskip

We develop further a new geometrical model of a discretized string,
proposed in [1] and establish its basic physical properties. The
model can be considered as the natural extension of the usual
Feynman amplitude of the random walks to random surfaces. Both
amplitudes coinside in the case , when the surface degenerates
into a single particle world line. We extend the model to open
surfaces as well. The boundary contribution is proportional
to the full length of the boundary and the coefficient of
proportionality can be treated as a hopping parameter of
the quarks. In the limit, when this parameter tends to infinity,
the theory is essentially simplified.
We prove that the contribution of a given
triangulation to the partition function is finite
and have found the explicit form for the upper bound.
The question of the convergence of the full partition
function remains open. In this model the
string tension may vanish at the critical point, if
the last one exists, and possess a
nontrivial scaling limit. The model contains hidden fermionic
variables and can be considered as an independent model of hadrons.\bigskip

\section{\it Introduction.} 

   It is well known, that lattice regularisation of gauge theories
can be
used to define the theory non-perturbatively [3]. In analogy with
this it
is important to find non-perturbative regularisation of the string
in order
to understand its complicated dynamical properties. At the same time
interacting random surfaces should play an important role in
describing
quantum fluctuations of the gauge systems at least in confining
phase [3].
The purpose of this paper is to develop and to study a geometrical
model of discretized string, which has been recently proposed in [1].

   Definition of the random surfaces model by dynamical
triangulation (DT)
allows to investigate many interesting models [2-6]. In this approach
continuum random surfaces are represented by triangulated surfaces.
Then in order to
describe statistical properties of two-dimensional surfaces, randomly
immersed into d-dimensional space $R^d$, it remains to choose a
suitable classical
action.
If the action is defined as the sum of the areas of the individual
triangles,
making up the surface- $A(S)$, then the partition function is ill
defined, because the area action does not suppress "spiky"
configurations [2,15]. If the action is proportional to the
length of all edges in a triangulation T,
that is the gaussian model with the action $A(L)$, then the string
tension does not scale at the critical point and the continuum limit
is also questionable [7]. Therefore it is not clear, whether there is
one or many universal classes of random surfaces, which are relevant
for the solution
of the problems, mentioned above.

   One of the possible conclusions, which can be done from these
results,
consists in the fact, that in order to reach a non-trivial continuum
limit
the classical action of the DT random surfaces must be chosen in the
way
to fulfil appropriate scaling behaviour. We shall call this dynamical
adjustment of the classical action to convenient scaling behaviour of
the model as {\it string fine tuning}.

  The question, which arises here, can be formulated in the following
way: is
there any simple principle, which allows to make a natural choice
of the classical action for DT surfaces, prior to solving the model?

   A new guiding principle was suggested in [1]. This {\it geometrical}
principle demands, that two surfaces, distinguished by a small
deformation
of the shape in the embedding space, must have close actions, that is,
these surfaces must have the same statistical weight.

   This principle is very restrictive and allows only few
possibilities. Let
us for example consider area action $A(S)$ and gaussian action $A(L)$.
In both cases the action is continuous with
respect to a small deformation of the DT surface, when we change the
coordinates of the vertexes of a given triangulation $T$. However, the
continuity of the action $A$ in the space of all triangulations
$\{T\}$ requires careful examination. Indeed, let us consider
two triangulations $T_{1}$ and $T_{2}$ with different numbers of
vertexes,
$N_{1}$ and $N_{2}$, but with the same Euler characteristic. There
always exists such a position of these vertexes,
that these two surfaces geometrically coincide.
Therefore the difference
$\delta A = A^{T_{1}} - A^{T_{2}} $ must tend to zero in
accordance with
our principle and this must hold good independently of the number of
vertexes. It is easy to check, that the area action $A(S)$
possesses the
defined property,
but the functional $A(L)$ does not [1]. Therefore the probability
distribution in the space of all triangulations $\{T\}$ is continuous
for
the action $A(S)$ and is discontinuous for the action $A(L)$.

   The question is, whether there are any new functionals
with the same properties, that is, they are
positive and  continuous in the space of all triangulations $\{T\}$ and
invariant under euclidean group of transformations. The answer is yes
[1].
Steiner functional [11] had these desirable properties, except
for the positivity.
But it can be extended in a such a way, that it will fulfil positivity
condition as well [1] (see below).

\begin{figure}
\centering
\includegraphics[width=6cm]{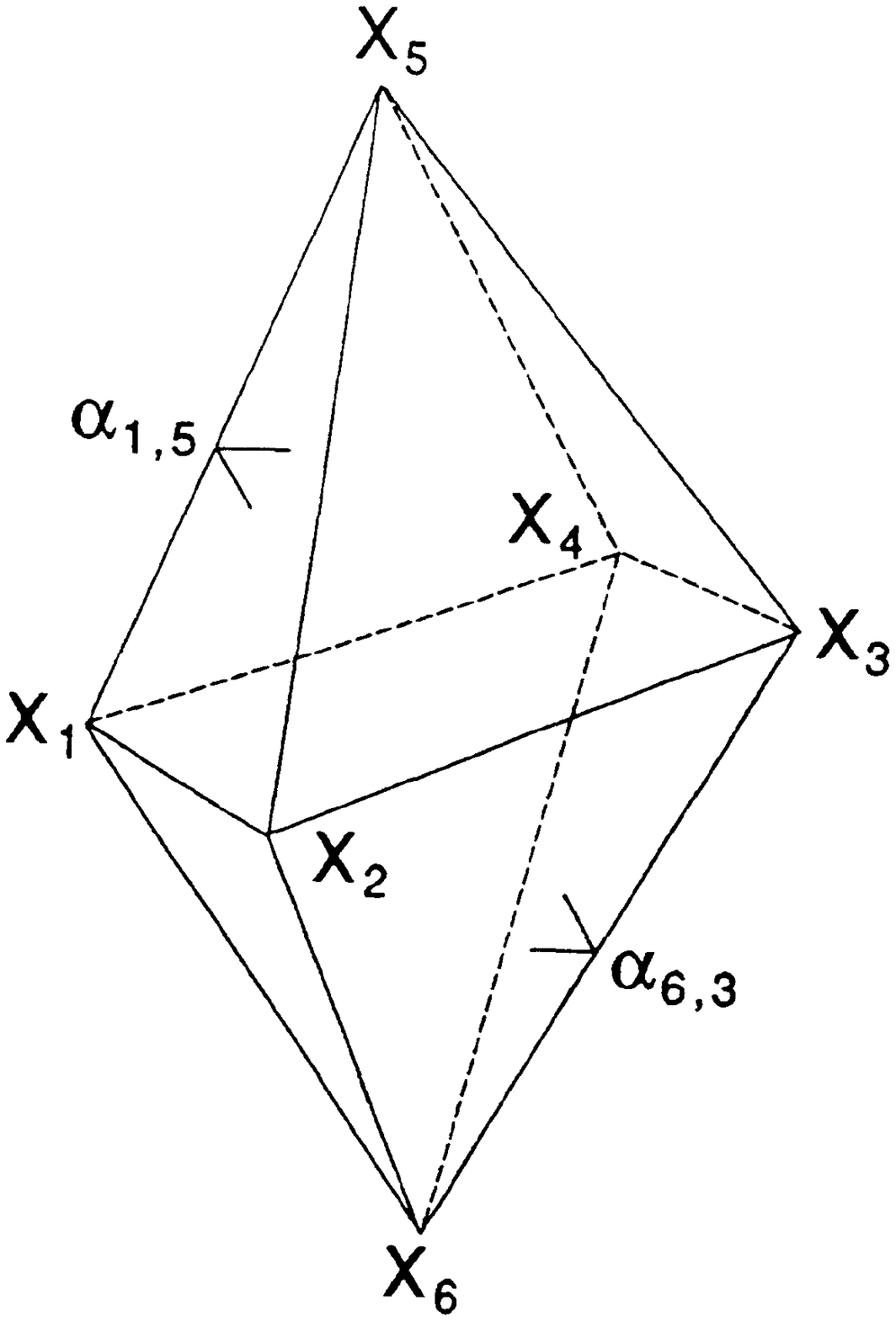}
\centering
\caption{The $A(M)$ is equal to $A(M) =1/2 \sum_{<i,j>}
\vert X_{i}-X_{j}\vert \cdot  \vert \pi - \alpha_{i,j} \vert$} .
\label{fig1}
\end{figure}

As a result we have an action, which is {\it modified} Steiner
functional [1]

$$A(M) = 1 / 2 \sum_{<i,j>} \vert X_{i}-X_{j}\vert \cdot
\Theta(\alpha_{i,j}),
\eqno(1)$$
where

   $$\Theta (\alpha) = \vert \pi - \alpha \vert\eqno(2)$$
and summation is over all edges $<X_{i},X_{j}>$ in triangulation T and
$\alpha_{i,j}$ is the angle between the embedded
neighbour triangles in $R^{d}$,
having a common edge $<X_{i},X_{j}>$ (see Fig.\ref{fig1}). This action indeed
has desirable properties.

   Thus DT surfaces, weighted by the area action $A(S)$ or by this
action $A(M)$, have continuous distribution of the statistical weights
in the space of all triangulations $\{T\}$. From other side,
a partition function of the DT surfaces is ill defined for the area action $A(S)$ and
converges for the gaussian action $A(L)$. The first question, which arises here,
is connected with the existence of the partition function for the theory
with the action $A(M)$. In this article we will prove, that the
contribution $Z_{T}$ of a given triangulation $T$ to the
partition function $Z$ is finite. The upper bound, which we
find out for these terms, does not permit us to state, that the
full partition function exists, but we expect, that it is
possible to improve this bound, because it is not the best.
Therefore this geometrical model still can be considered
as possible
variety of discretised string.

 At the beginning we shall present simple qualitative arguments,
demonstrating the convergence of the partition
function $Z$ [1]. The partition function of our surface model is

$$Z(\beta) = \sum_{over surfaces} e^{-\beta \cdot A(M)}
           =  \sum_{T \in \{ T \}} Z_{T} (\beta) .$$
   The first argument is based on Minkowski inequality [10]  (see also
[9,12]),  which tells, that on arbitrary surface the functional $A(M)$
is always greater, than the square root of the surface area

                       $$A^2 (M) \ge 4\pi S,\eqno(3a)$$
and equality takes place only for sphere, that is, among all surfaces
with
fixed area, the sphere has minimal $A(M)$. From (3) one can conclude,
that the convergence of the partition function is in any case better
than for the area action, because
$$e^{-A} \le e^{-\sqrt {4\pi S}}, \eqno(3b)$$
and that the maximal distribution carries the
surfaces, close to sphere. This fact has very
elegant consequence: quasi-classical expansion, when $\beta \rightarrow
\infty $, must be done
around sphere. In that regime there is no big difference between area
action and $A(M)$ action. This is especially valuable, because strong
coupling expansion of the gauge theory on a lattice is described
by the area action [3].

 As we already explained, the action (1) is not exactly the
Steiner functional, and inequality (3) was established
only for convex surfaces.
Initially the Steiner functional was written for convex
surfaces and for them  $\Theta$ is equal to $\pi - \alpha$ [11].
For our
purposes we need a certain extension of the classical Steiner
functional
to all surfaces. The extension (1) is made up in a such a way,
that the
action is always positive and concave surfaces with the same area
have
larger action than the convex ones. Therefore, for the modified
functional $A(M)$ inequality (3)
takes place for all surfaces. The inequality (3) is very important
for
the physical interpretation of $A(M)$, because it establishes an
absolute
minimum of the functional $A(M)$ and guarantees, that the surfaces
will
not collapse with large probability to crumple surfaces. Note,
that this extension also provides $\it locality$ of the classical
action [1].

The next qualitative argument, displaying the convergence of the
partition function, consists in the observation, that the Steiner
functional for the convex surfaces geometrically is the $\it mean$
$\it size$ of the surface [12-14]

   $$A(M) = \int L_{g} \ dg,\eqno(4)$$
where $L_{g}$ is the length of the orthogonal projection of the
surface T to the line $g$, which crosses the origin in $R^d$. $dg$
is the
invariant measure of these lines [14-16] . For example, if $T$
is sphere,
then
$L_{g} = 2R$ and $A(M) = 4 \pi R$ . It is easy to see from (4),
that $A(M)$ increases with the diameter $\Delta$
of the
surface
(here $\Delta$ is the largest distance between two points on the surface),
so
the statistical weight of the surface decreases.
For example, spiky configurations are strongly
suppressed, because they have large size. In the case of area
action, the action
does not always increase with the diameter of the surface and the
corresponding weights can even be constant.

   In the next section we rigorously define the surface action
$A(M)$ (1) [1].
We present the necessary geometrical properties of this functional
and consider few examples. In fact, if the surface degenerates into
single curve, the action $A(M)$ becomes proportional to a full length
of the curve and transition amplitude coincides with the one for the
Feynman path integral. This means that $A
(M)$ is the natural extension of usual Feynman amplitude
of the random walks to random surfaces.

If the surface has boundaries,
"created by virtual quarks or by external sources", then the boundary
part of the action is proportional to the full length of the boundary.
The coefficient of proportionality can in principle be changed
(see (2b)) and treated as the hopping-parameter
of the quarks. The limit, in which this parameter tends to infinity,
corresponds to pure QCD, otherwise we would have dynamical quarks.
Here we use QCD terminology merely because random surfaces should
describe the fluctuation of gauge degrees of freedom at least in confining
phase, but certainly these results do not depend on the use of this analogy.
In same sense one can consider this model as an independent theory.

   In the third section we prove that the contribution
$Z_{T}$ of a given triangulation $T$ to the partition
function $Z$ is finite. At the begining we will find the
following bound
$$Z_{T} \leq (Const \cdot 2^{\vert T \vert /2} )^{d (\vert T \vert -1)}$$
where $\vert T \vert $ is the number of vertexes and $d$ is the
dimention of the space. Then, using "shape" coordinates [15],
we improve this bound and find, that

$$Z_{T} \leq (Const)^{d (\vert T \vert -1)} \cdot (d(\vert T
\vert - 1 ))!$$
This bound does not permit to prove the convergence of
the full partition function, but it is important to note, that
the volume of the "shape"
coordinates was crudely estimated. Therefore we expect
that it is possible to improve this bound.
This section is more technical, but
permits to clarify contents of the theory, particularly geometrical
meaning of the extension (1).

   The fourth section is devoted to the discussion
of the critical behaviour of the model and we will show that the
theorem of
Ambjorn and Durhuus [7], concerning the non-vanishing of the string
tension
of the gaussian model, is not valid in our case, so the string
tension may tend to zero at the critical point. In the last section
we
discuss possible extensions of this model and compare them with
the gaussian one.

\section{\it Geometrical and physical properties
of $A(M)$}

Let us consider closed surfaces in $R^d$, given by the mapping of the
vertices of some fixed triangulation T into the euclidean space
$R^d$ [2].
Triangulations T are defined as connected, two-dimensional abstract
simplical  complexes with fixed topology [2]. The surfaces to be identified with
a piecewise linear surfaces embedded into $R^d$ (see fig.1).
The coordinates of
the vertexes in $R^d$ are denoted as $X_{i}^{\mu}$,
where $\mu = 1,..,d$  and $i = 1,..,\vert T \vert$ . $\vert T \vert$
is the
number of vertexes on T. We shall use the same letter T to denote the
surface.
The action is given by [1]

$$A(M) =1/2 \sum_{<i,j>} \vert X_{i} - X_{j}\vert \cdot \Theta
(\alpha_{i,j}),\eqno(1)$$
where

           $$\Theta(\alpha) = \vert \pi - \alpha \vert\eqno(2)$$
and summation is over all edges $<X_{i},X_{j}>$ ( i and j are
the nearest neighbours in T ), $\alpha_{i,j}$ is the angle between the
embedded
neighbour triangles in $R^d$ having a common edge $<X_{i},X_{j}>$.
This expression essentially differs from the gaussian action $A(L)$,
because here all lengths of the edges are directly multiplied by the
corresponding angles $\Theta(\alpha_{i,j})$ (see fig.1). This means
that the
edges are weighted in a more complicated way compared to a gaussian
model:
sometimes they contribute in full,$\alpha \approx 0,2 \pi$, sometimes
they do
not contribute at all, $\alpha \approx \pi$. This circumstance provides
the
action $A(M)$ by peculiar geometrical and physical properties.

As it was already explained, the module in (2) corresponds to a
specific
extension of the Steiner functional to all surfaces. The extension (1)
coincides
with the Steiner functional for convex surfaces and therefore inherits
the same
geometrical nature. At the same time, extension (1) provides
relative suppression
of the crumple surfaces, because for them $\alpha$ is close to zero
or to
$2\pi$, so $\Theta$ is in its maximum. Finally we have locality of
the classical action.

 To have some intuitive experience let us consider few examples.
The action
$A(M)$ has the dimension of the length and expresses, as it follows
from the definition (1) and representation (4),
the mean size of the surface. Therefore even when the
surface
degenerates into a thin tube, the action does not vanish, as it
happens in
the case of area action, because this tube has non zero size. If the
surface
degenerates into a single curve, then it is easy to see from the
definition (1), that the action is proportional to the full length $L$
of that curve

                $$A(M) = L/2 \cdot \pi .\eqno(5a)$$

So $A(M)$ correctly weights degenerated surfaces, it is proportional
to the full length of the curve, and the transition function coincides
with
the one for the Feynman path integral.This means that $A(M)$ is the
natural
extention of usual Feynman amplitude of the random walks to random
surfaces.
The area action is unable to describe such surfaces and the gaussian
action
"over-count" these configurations.This leads to spikes,
growing out of the surface in the model with area action, and
to non-scaling behaviour of the string tension in the gaussian
model [2,7].

 The coefficient of proportionality on the right-hand side of (5a) coded
the information about the way the surface squeezed into the curve. For example,
if the surface collapsed and crumpled, then this coefficient increases and
is equal to the number of crinkles

       $$A(M) = L/2 \cdot k \pi,\eqno(5b)$$
where $k$ is the number of crinkles.

Up to now we have considered only closed surfaces, but the definition (1)
has
natural meaning for the {\it open} surfaces as well. It is reasonable to
take $\alpha_{i,j}$ on the boundary edges equal to zero or to $2\pi$. Then
from the definition (1) it follows, that the boundary part of the action
is proportional to the full length of the boundary

        $$A(M)_{boundary contribution} = L/2 \cdot \pi,\eqno(6a)$$
where $L$ is the full length of the boundary. This fact has important
physical consequence: the quark loop amplitude is proportional to the
length of their world line. Let us consider an extension of the action (1),
in which $\Theta$ satisfies to more general conditions [1]

  $$\Theta(2\pi - \alpha) = \Theta(\alpha),$$
  $$\Theta(\pi) = 0,$$
  $$\Theta(\alpha) \geq 0,\eqno(2a)$$
then we will get

     $$A(M)_{boundary  contribution} = L/2 \cdot \Theta(0).\eqno(6b)$$
So the full length of the boundary is multiplied by $\Theta(0)$.
$\Theta(0)$ plays the role of the hopping-parameter of the quarks.
In fact, if we take $\Theta (0)$ to be very large, then the probability
of the quark loop creation tends to zero (see (6b) and one can treat
$\Theta (0)$
as the logarythm of the quark masses. At the same time, in the limit
$\Theta(0) \rightarrow \infty$ the string tension may
have finite limit, because,
as we will see in the next section, string tention is formed by
fluctuations of the surface near $\alpha \approx \pi, \Theta(\pi)
\approx 0$.
These  fluctuations do not disappear even if $\Theta(0) = \infty$
(see (2b).
If so, then this limit corresponds to pure $QCD$, that is, quarks
are not dynamical. When $\Theta(0) \rightarrow \pi$, then we will have
dynamical quarks.

   At the end of this section we would like to emphasise, that
unless the action $A(M)$ has many different representations,
we consider the expression (1) as more fundamental, because it
is well defined for extremely large class
of surfaces, including degenerate surfaces. The model is well
defined
in any dimension and for arbitrary topologies.
Of course it coincides with other  expressions in special cases.

It is also important to note,
that this model can be considered as a sort of a theory
with extrinsic curvature action [19,20,1].
 \medskip

\section{\it The partition function}

 Partition function is defined usually as in [2]

$$Z(\beta) = \sum_{T \in \{ T \}} \rho (T) \int e^{-\beta A(M)}
\prod_{i \in T}
dX_{i} ,\eqno(7)$$
where $\{ T \}$ denotes some set of closed triangulations T , $\rho (T)$
are their weights, $dX_{i}$ are the measures in $R^d$.
One vertex $X_{f}^{\star}$ is fixed to
remove translation invariant zero-mode and we should specify the summation
weight $\rho (T)$ in the definition (7).This can
be done as a consequence of our first geometrical principle. It demands,
that the distribution of the statistical weights must be continuous in
the space of all triangulations $\{T\}$.To satisfy this condition we must
choose  $\rho
(T) = 1$ .

   Now we would like to prove, that the contribution of a given
triangulation $T$ to the partition function is finite
and would like to find the explicit form of the upper bound.
With this aim, in our first approach we will
successively perform the integration over all vertexes in T.

For a while let us denote the coordinates of the arbitrarily
chosen vertex $X_{i}$
by $X$ and its nearest neighbour vertexes by  $X_{1},..,X_{q}$. They are
ordered cyclically around $X$ and $q$ is the order of $X$ (the number of
ingoing edges). The vertex of the triangle, which lies opposite to
the triangle
$<X,X_{i},X_{i+1}>$ and has a common edge $<X_{i},X_{i+1}>$, we will
denote by $Y_{j}$, where $j=1,..,p$ and $1\leq p \leq q$ (see fig.2).

\begin{figure}
\centering
\includegraphics[width=6cm]{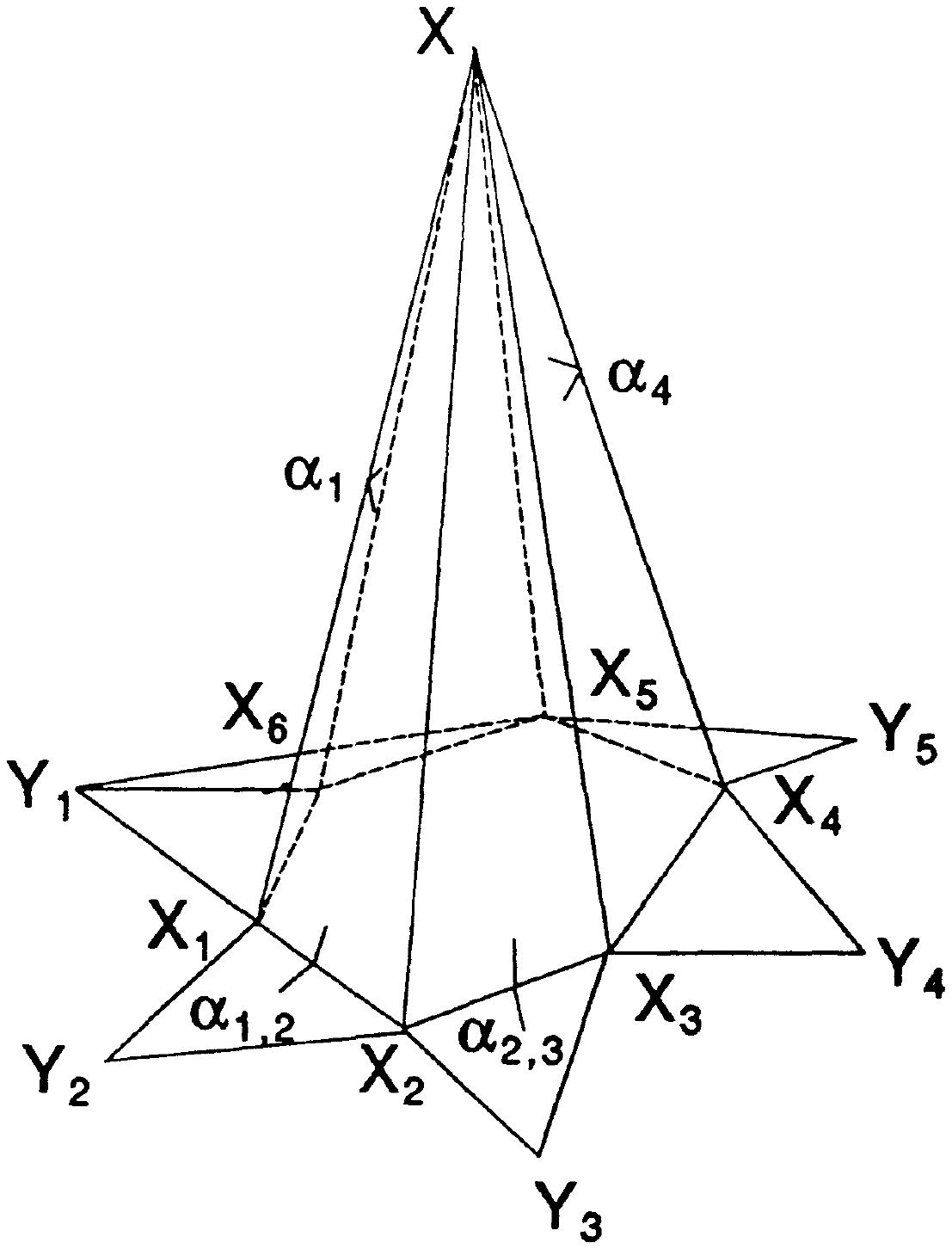}
\centering
\caption{The $A(M)$ depends on $X$ through the length of the edges
$\vert X-X_{i}\vert$ and through the compact angle variables
$\alpha_{i}$ and $\alpha_{i,i+1}$ .}
\label{fig2}
\end{figure}

The part of the classical action $A(M)$ (1), which depends on $X$, has
   the form:

            $$A(M) = A_{x} + A_{xy} + A,\eqno(8)$$

$$A_{x} = \sum_{i=1}^{q} \vert X - X_{i} \vert \cdot
\Theta(\alpha_{i}),\eqno(9)$$

$$A_{xy} = \sum_{i=1}^{q} \vert X_{i} - X_{i+1} \vert \cdot
\Theta(\alpha_{i,i+1}),\eqno(10)$$
where

$$\Theta(\alpha_{i}) = \vert \pi - \alpha_{i} \vert ,\eqno(11)$$

$$\Theta(\alpha_{i,i+1}) = \vert \pi - \alpha_{i,i+1} \vert \eqno(12)$$
and $X_{q+1} = X_{1}$. The $\alpha_{i}$ is the angle between two
triangles,
having common edge  $<X,X_{i}>$  and $\alpha_{i,i+1}$  is the one
with a common
edge $<X_{i},X_{i+1}>$ (see Fig.\ref{fig2}). This decomposition is very natural,
because
$A_{x}$  depends on $X$ through the length of the edges
$\vert X - X_{i} \vert$ and through the corresponding compact angle
variables
$\alpha_{i}$. As far as $A_{xy}$  is concerned, it depends on $X$
{\it only} through the compact
variables $\alpha_{i,i+1}$. The integral over $X$ has the form

$$Z(X_{i},Y_{j}) = \int e^{-A_{x} - A_{xy}} dX ,\eqno(13)$$
where

    $$e^{-A_{x}} \equiv g(X,X_{i}) ,$$
    $$e^{-A_{xy}} \equiv f(X,X_{i},Y_{j}). \eqno(14)$$
It seems reasonable to use the bound  $e^{-A{xy}} \leq 1$ in (13)
and then to
prove the convergence of the remaining integral (see (16). But in
that case
the next integration over the nearest neighbour vertex $X_{i}$ will
create
difficulties, because $A_{xy}$ has few terms of the action with
edge angles $\alpha_{i,i+1}$, which also belong to a vertex
$X_{i}$ . These terms will be absolutely nessesary, when we will perform
integration over $X_{i}$, because they allow to bind the sum of all
edge angles, belonging
to $X_{i}$, from below (see (19),(20)). So, we will
treat the term $e^{-A_{xy}}$ in a
more gentle way, that is we would like to find following expression
for $Z$ after
integration over $X$ in (13),
$$Z(X_{i},Y_{j})= Const \cdot e^{-A_{x^{*}y}},$$
where $X^{*}$ is an "average" position of the vertex $X$.
Following inequality takes place:

$$m\cdot e^{-A_{x}} \leq e^{-A{x} - A_{xy}} \leq M \cdot e^{-A_{x}}
,\eqno(15)$$
in which  $M(X_{i},Y_{j})$  and $m(X_{i},Y_{j})$  are the maximum and
minimum
of the function $f(X,X_{i},Y_{j})$ over $X$ at fixed $X_{i}$ and $Y_{j}$.
The function  $f(X,X_{i},Y_{j})$
continuously
depends on $X$ through the compact angle variables $\alpha_{i,i+1}$,
therefore the maximum and minimum actually exist (Weierstrass theorem).
Now let us prove the convergence of the integral

    $$Z(X_{i}) = \int e^{-A_{x}} dX .\eqno(16)$$

Let denote by $r$ the minimal radius of the sphere, which contains all
vertexes
$X_{1},..,X_{q}$ and choose the origin in the centre of that sphere.
Then we will have

$$Z(X_{i}) = \int_{\vert X \vert \leq 2r} e^{-A_{x}} dX +
\int_{\vert X \vert \geq 2r} e^{-A_{x}} dX .\eqno(17)$$
Inside sphere, in the first integral, one can bound the integrant by one

$$\int_{\vert X \vert \leq 2r} e^{-A_{x}} dX \leq
{\pi^{d/2} \over \Gamma(1+d/2)} \cdot (2r)^{d} \eqno(18)$$
and change every edge length $\vert X - X_{i} \vert$ by smaller quantity
 $\vert X \vert - r$ in the second integral

$$\int_{\vert X \vert \geq 2r} e^{-A_{x}} dX \leq
\int_{\vert X \vert \geq 2r} e^{-(\vert X \vert - r) \sum_{i=1}^{q}
\Theta(\alpha_{i})} dX .\eqno(19)$$
We must bound the sum of all edge angles, belonging to vertex $X$, from
below.
The desirable answer is

   $$0 < \Theta_{min} \leq \sum_{i=1}^{q} \Theta(\alpha_{i})
\leq q \pi.\eqno(20a)$$
The proof of this fact will be presented in separate place. It is
important
to note, that this number $\Theta_{min}$ does not depend on $r$ and $q$,
but
depends on proportion $k$, by which the integral (16) was divided into
two
parts (see (17)).For $d=3$ we get

$$\Theta_{min} = 2\pi \cdot \sqrt {(1-k^{-2})},\eqno(20b)$$
and if $k = 2$, as it was in (17),
$$\Theta_{min} = \pi \sqrt {3}. \eqno(20c)$$

Inserting (20a) in (19) we get

$$\int_{\vert X \vert \geq 2r} e^{-(\vert X \vert - r) \cdot
\Theta_{min}} dX
= $$ $${2\pi^{d/2} \over \Gamma(d/2)} \cdot
e^{\Theta_{min} r} \cdot \int_{2r}^\infty e^{-\vert X \vert
\Theta_{min}}
(\vert X \vert )^{d-1}\,dX \leq \left( Const \over \Theta_{min}
\right)^{d}.
\eqno(21)$$
The radius $r$ is always smaller than the sum of all distances between
$X_{1},..,X_{q}$

   $$r \leq \sum_{i=1}^{q} \vert X_{i} - X_{i+1} \vert , \eqno(22)$$
thus

$$Z(X_{i}) \leq \left[ Const\sum_{i=1}^{q} \vert X_{i} - X_{i+1} \vert
\right]^{d}
\eqno(23)$$
It is helpful to compare this bound with the same quantity for the
model with area action [2]

$$Z(X_{i}) = \int e^{-A_{x}(S)} dX \geq 1 / {\left[ Const \sum_{i=1}^{q}
(X_{i} - X_{i+1})^{2} \right]^{d/2}}. \eqno(24)$$
The difference consists in the fact, that in our case the weight
function is equal to one in the region of order $r$, while for the model
with
area action the same region is of order $1/r$.

   Because the last integral (23),(16) is finite, one can integrate
inequality (15) over $X$

$$m \cdot Z(X_{i}) \leq Z(X_{i},Y_{j}) \leq M \cdot Z(X_{i}), \eqno(25)$$
or

$$m \leq {Z(X_{i},Y_{j}) \over Z(X_{i})} \leq M . \eqno(26)$$
This relation is a direct analog of the classical formula from calculus

$$m \leq {\int_{-\infty}^\infty f(X)g(X) dX \over \int_{-\infty}^\infty
g(X) dX } \leq M,$$
where $g(X)$ is an integrable function in $[-\infty ,\infty ]$ and
$m \leq f(X) \leq M$, so

$$\int_{-\infty}^\infty f(X)g(X) dX = f(X^{\star}) \int_{-\infty}^\infty
g(X) dX. $$
The $f(X,X_{i},Y_{j})$  is continuous function of $X$ and takes all its
values
between  $m(X_{i},Y_{j})$  and  $M(X_{i},Y_{j})$  (Bolzano-Cauchy
theorem),
therefore there exists such a position of the vertex $X = X^{\star}$, for
which

$$Z(X_{i},Y_{j}) = f(X^{\star},X_{i},Y_{j}) \cdot Z(X_{i}), \eqno(27a)$$
or

$$Z(X_{i},Y_{j}) = e^{-\sum_{i=1}^{q} \vert X_{i} - X_{i+1} \vert \cdot
\Theta(\alpha_{i,i+1}^{\star})} \cdot Z(X_{i}). \eqno(27b)$$
Using (23) we get

$$Z(X_{i},Y_{j}) \leq e^{-\sum_{i=1}^{q} \vert X_{i} - X_{i+1} \vert \cdot
\Theta(\alpha_{i,i+1}^{\star})} \cdot \left[ Const\sum_{i=1}^{q}
\vert X_{i} - X_{i+1} \vert  \right]^{d}, \eqno(28)$$
where

$$\alpha_{i,i+1}^{\star} = \alpha_{i,i+1}(<X^{\star},X_{i},X_{i+1}>,
<X_{i},X_{i+1},Y_{i}>) \eqno(29)$$
So the quantity $Z(X_{i},Y_{j})$ is finite and we preserve the part of
the action, which contains the edge angles $\alpha_{i,i+1}$.
Now they can
"participate" in the subsequent integration over all $X_{i}$ and
permit to
bound the sum of all edge angles, belonging to $X_{i}$, from below,
as it was
already done for the vertex $X$ (19-21).

The main trouble
consists in the fact, that the vertex $X^{\star}$ can in principle
be settled down on infinity,
because in that point $f(X,X_{i},Y_{j})=e^{-A_{xy}}$ is non zero and
finite.
The last circumstance insists on us to change slightly the strategy.
We shall
use another decomposition in (13)

  $$Z(X_{i},Y_{j}) = \int h^2 dX ,\eqno(30)$$
where

      $$h = e^{-(A_{x} + A_{xy})/2}. \eqno(31)$$
Repeating calculations for the last decomposition, we will get

 $$Z(X_{i},Y_{j}) = h^{\star} \cdot \int h dX \leq Const \cdot
 h^{\star},\eqno(32)$$
where
$$h^{\star} = e^{-(A_{x^{\star}} + A_{x^{\star}y})/2} =
e^{-1/2\sum_{i=1}^{q} \vert X^{\star} - X_{i} \vert \cdot
\Theta(\alpha_{i}^{\star}) + \vert X_{i} - X_{i+1} \vert
\cdot \Theta(\alpha_{i,i+1}^{\star})} \eqno(33)$$
and
$$\alpha_{i}^{\star} = \alpha_{i}(<X^{\star},X_{i},X_{i-1}>,
<X^{\star},X_{i},Y_{i+1}>). \eqno(34)$$
Now the vertex $X^{\star}$ is sited on a finite distance from the
vertexes
$X_{i}$ and $Y_{j}$. In fact, the function $h(X,X_{i},Y_{j}) =
e^{(-A_{x} - A_{xy})/2}$ reaches the
minimum at infinity, because at that point  $\vert X^{\star} \vert =
\infty$, the
value of the $A_{x^{\star}} = \infty$ and  thus $h^{\star}= 0$. But
the value of full integral (32) is strictly positive, therefore
the "average" vertex $X^{\star}$ cannot be at infinity (see (33)).

The final result of the integration over the vertex $X$ consists in
three facts:\\
\\
   i)now two of the initial vertexes are fixed $X_{f}^{\star}$ and
   $X^{\star}$, but the position of the average vertex $X^{\star}$
   depends on its neighbour vertexes $X_{i}$ and $Y_{j}$ ,\\
\\
   ii)the effective classical action does not change its own
      geometrical nature,that is all terms, belonging to the
      vertex $X^{\star}$, still remain (33),\\
\\
   iii)the effective action decreases by the factor two (33).\\
\\
From (32) we get

$$\int e^{-A(M)} \prod_{i \in T \setminus X_{f}^{\star}} dX_{i} = $$
$$\int e^{-(A_{x^{\star}} + A_{x^{\star}y})/2} \cdot e^{-A - (A_{x} +
A_{xy})/2}
\cdot \prod_{i \in T \setminus X_{f}^{\star}} dX_{i} \leq $$
$$Const \cdot \int e^{-(A_{x^{\star}} + A_{x^{\star}y})/2} \cdot e^{-A}
\prod_{i \in T \setminus X_{f}^{\star},X^{\star}} dX_{i} \leq $$
$$Const \cdot \int e^{-(A +A_{x^{\star}} + A_{x^{\star}y})/2}
\prod_{i \in T \setminus X_{f}^{\star},X^{\star}} dX_{i},
\eqno(35)$$

Now it remains to integrate over the rest vertexes. Below we will
assume, that the vertex  $X^{\star}$ depends on $X_{i}$ and $Y_{j}$
in a such way, that it is always sited within the region  of the
sphere of radius $r$ (22), which contains all vertexes
$X_{1},..,X_{q}$. Then every subsequent integration will change the
exponent by the factor $1/2$, so finally we get

$$Z_{T} \leq (Const \cdot 2^{\vert T \vert /2})^
{d( \vert T \vert - 1)}
\cdot e^{-A(M^{\star})/
{2^{\vert T \vert-1}}},\eqno(36)$$
where all vertexes in $A(M^{\star})$ are "imaginary" and are situated
on a finite distance from each other. The constant on the right-hand
side of (36) depends only on d .

{}From (36) and (7) it follows, that

$$Z(\beta) \leq \sum_{T \in \{ T \}}
(Const \cdot 2^{\vert T \vert /2}/ \beta)^{d( \vert T \vert - 1)},$$
but this bound is not sufficient to
prove the convergence of the full partition function. It is
useful to compare this bound with the one in the gaussian model [19].

Let us
return to our assumption of the "week" dependence of the vervex
$X^{\star}$ from its neighbours. To control this dependence
we must take into account the fact that
the full action $A(M)$ is the mean
size of the surface and therefore $X^{\star}$ can not "run" far
away from the origin (which we will take in the
fixed point $X_{f}^{\star}$).
Explicit estimates are very fragile, because we must use a local
and global properties of $A(M)$ at the same time.

In this place we came to a point, that the convergence can be
proved only by using global character of $A(M)$. Indeed

         $$A(M) \geq \Delta ,$$
where $\Delta$ is the diameter of the surface (largest distance
between two points on the surface) and let us define $R$

$$R = \sqrt{X^{2}_{1}+....+X^{2}_{\vert T \vert -1}} \leq
\sqrt{(\vert T \vert -1)} \cdot \Delta ,$$
where the origin is in the point $X_{f}^{\star}$. So we have

$$A(M) \geq R/\sqrt{(\vert T \vert -1)}$$
and therefore for $Z_{T}$ we will get

$$\int e^{-A(M)} \prod_{i \in T \setminus X_{f}^{\star}} dX_{i} \leq $$
$${2\pi^{d(\vert T \vert -1)/2} \over \Gamma(d(\vert T \vert -1)/2)}
\cdot
\int_{0}^\infty e^{-R/\sqrt{\vert T \vert -1}}
R^{d(\vert T \vert -1)-1}\,dR =$$
$${2\pi^{d(\vert T \vert -1)/2} \over
\Gamma(d(\vert T \vert -1)/2)} \cdot
(\vert T \vert -1)^{d(\vert T \vert -1)/2} \cdot
{(d(\vert T \vert -1))!\over d(\vert T \vert -1)} \eqno(35a)$$

  The third possibility is to introduce a new coordinate system
in which $A(M)$ is an independent variable. A remaning part of
the variables we will take as angles $\vec{\alpha}$ or "shape"
variables [15] (they don't coincide with $\alpha_{i,j}$) ,
so we have

$$\int  \prod_{i \in T \setminus X_{f}^{\star}} dX_{i} =
\int A(M)^{d(\vert T \vert -1)-1} \cdot J(\vec{\alpha})\, dM(A)\,
d\vec{\alpha} ,$$
where $J(\vec{\alpha})$ is the Jacobean of this transformation.
Integrating both sides over the $X's$ in $R^{d}$ with the
condition, that for them
$A(M) \leq A$, we will get

$$\int J(\vec{\alpha})\,d\vec{\alpha} =
\int_{\leq A} \prod_{i \in T
\setminus X_{f}^{\star}} \, dX_{i}  /  \int_{0}^{A}
A(M)^{d(\vert T \vert -1)-1}\, dM(A) .$$
Because every coordinates $X's$ are bounded at list in the region
$0 \leq \vert X_{i} \vert \leq A$, it follows, that

$$\int J(\vec{\alpha})\,d \vec{\alpha} \leq
d(\vert T \vert -1) \cdot \left ({\pi^{d/2} \over \Gamma(1+d/2)}\right )^
{\vert T \vert -1}. $$
Therefore for full integral we will get

$$\int e^{-A(M)} \prod_{i \in T \setminus X_{f}^{\star}} dX_{i} =  $$
$$\int e^{-A(M)} A(M)^{d(\vert T \vert -1)-1} \cdot J(\vec{\alpha})\,
dM(A)\,d\vec{\alpha} \leq
(d(\vert T \vert -1))!\left ({\pi^{d/2} \over \Gamma(1+d/2)}\right )^
{\vert T \vert -1} \eqno(35b). $$
By the same technic one can prove, that the contribution of a
fixed triangulation to the loop Green functions are also
finite, but this bound still leave the question of the full
convergence open. \medskip

\section{\it Loop Green functions}

   The loop Green functions can be defined as usually [2]:

$$G_{\beta}(\gamma_{1}...,\gamma_{n}) = \sum_{T \in \{ T \}} \rho (T)
\int e^{-\beta A(M)} \prod_{i \in T \setminus \partial T} dX_{i}
\eqno(37)$$
where $\gamma_{1}...,\gamma_{n}$ are closed polygonal loops in $R^{d}$
with
$k_{1}..,k_{n}$ corners, and $\{ T \}$ denotes the set of triangulations
with
$n$ boundary components.

If the loop Green functions are finite for
sufficiently
large $\beta$, then one can define string tension    [2]

$$\sigma(\beta) = -\lim_{R,L \rightarrow \infty} \ln G_{\beta}
(\gamma_{R,L})
/{RL} , \eqno(38)$$
where $\gamma_{R,L}$ is rectangle with sides of length $R$ and $L$.
Ambjorn and Durhuus have proved, that string tension of the gaussian
model does not vanish at the critical point, because the minimal
surface
dominates in the functional integral.Quantum fluctuation does not
contribute and cannot lower the "classical" string tension [7].
They found, that [7]

$$G_{\beta}(\gamma_{R,L}) \leq e^{-2\beta RL} \cdot
G_{\beta}(O_{R,L}).\eqno(39)$$
The $G_{\beta}(O_{R,L})$ is the loop Green function,
where the loop  $\gamma_{R,T}$ is contracted into one point  $O$.
For this Green function they found the bound [7]

$$G_{\beta}(O_{R,L}) \leq e^{C(\beta) \cdot (R+L)},\eqno(40)$$
where $C(\beta)$ is finite for $\beta > \beta_{c} > 0$, thus [7]

$$\sigma (\beta) \geq 2\beta .\eqno(41)$$
and string tension does not vanish at $\beta_{c}$.

In our case we have, that

$$A(M_{R,L}) = A_{min}(M_{R,L}) + A(M,O_{R,L}), \eqno(42)$$
where $A_{min}(M_{R,T})$ is the minimum of $A$ for the surfaces with
rectangular boundary
$\gamma_{R,L}$ and $A(M,O_{R,L})$ is the remaining part of the action.
It is
easy to see, that

 $$A_{min}(M_{R,L}) = 2(R+L)\cdot \pi . \eqno(43)$$
Inserting (42-43) in (37), we get

$$G_{\beta}(\gamma_{R,L}) = e^{-\beta (R+L) \cdot \pi} \cdot
G_{\beta}(O_{R,L}),\eqno(44)$$
so the minimal surface does not contribute at all (compare with (39)).
Only quantum fluctuations are important:

$$\sigma(\beta) = -\lim_{R,L \rightarrow \infty} \ln G_{\beta}(O_{R,L})
/{RL}.$$

For the general model (2b) we will get

$$G_{\beta}(\gamma_{R,L}) = e^{-\beta (R+L) \cdot \Theta(0)} \cdot
G_{\beta}(O_{R,L}). \eqno(45)$$

  Let us consider the case, when $\Theta (0) \to \infty$.
In this limit
the theory is essentialy simplified. In fact, all fluctuations,
in which
the surface considerably crumpled, that is $\alpha$ is close to zero
or to
$2\pi$ and $e^{-A(M)} \to 0$, are strongly suppressed. Nevertheless,
in this limit the surface with the boundary $\gamma_{R,L}$ can
fluctuate
near the angles $\alpha \approx \pi$, where $\Theta (\alpha)$
is close to zero.
Now one can estimate contribution of such fluctuations to the loop
Green function

$$G_{\beta}^{T}(\gamma_{R,L}) \approx a^{T} \cdot
e^{-\sigma_{T}(\beta) \cdot RL}. $$
The difference with the gaussian model consists in slower
decrease of the
series expansion (37). To convince of that let us use the inequality
$$A(M) \leq (\sum_{<i,j>} \vert X_{i}-X_{j}\vert) \cdot
(\sum_{<i,j>} \Theta(\alpha_{i,j})) \leq
\sum_{<i,j>} \vert X_{i}-X_{j}\vert \cdot 3\pi
(\vert T \vert - Euler \cdot charac.),\eqno(3c) $$
so for torus
$$e^{-3\pi \vert T \vert \cdot A^{T}(L)} \leq  e^{-A^{T}(M)}
\eqno(3d).$$
Therefore the string tension may have nontrivial scaling limit in
this theory.

\medskip

\section{\it Discussion} 

It is useful to compare (1) with the gaussian model. For that let us
consider the action

$$A = \sum_{<i,j>} (X_{i} - X_{j})^{2} (1 + g\cos(\alpha_{i,j})),
\eqno(46)$$
where $g$ is the coupling constant $0 \leq g \leq 1$. If $g = 0$,
then this model coincides with gaussian one and, if $g = 1$, it
coinsides
with (1),
(2b), where

$$\Theta(\alpha) = 1 + \cos(\alpha),\eqno(47)$$
just in (46) we have the squares of the lengths.

When $g\ne 1$, the edges contribute with the weights, variating
in the region $[(1-g)(X_{i} - X_{j})^2 ; (1+g)(X_{i} - X_{j})^2]$.
These weights do not vanish for all values of the angle
$\alpha_{i,j}$,

$$e^{-(1+g)\sum_{<i,j>}(X_{i} - X_{j})^2} \leq e^{-A} \leq
e^{-(1-g)\sum_{<i,j>}(X_{i} - X_{j})^2} \eqno(48)$$
The critical temperatures for these boundary models are equal to
$\beta_{c}/
(1+g)$ and $\beta_{c}/(1-g)$ respectively.The critical exponents are
the same
ones.

For the most interesting point $g=1$ the weight function
can vanish, $\Theta(\pi)
= 0$ (see (47),(46)), and does not permit us to find the bound.
Exactly the same property ensures the continuity of the $A(M)$ in the
space
of all triangulations [1], that is why the model was examined
separatelly.

The model, in which boundary contribution can variate (see (6b)) ,
has the form

$$A = \sum_{<i,j>} (X_{i} - X_{j})^{2} (1 + \cos(\alpha_{i,j}))/
{(1 - g\cos(\alpha_{i,j}))},\eqno(49)$$
where

$$\Theta(\alpha) = 1+\cos(\alpha)/{1-g\cos(\alpha)}.$$
The quark loop amplitude is proportional to

$$e^{-\beta \cdot \Theta(0) \cdot L},\eqno(50)$$
where

$$\Theta(0) = 2/(1 - g).\eqno(51)$$
Pure $QCD$ corresponds to $g=1$. The last two models can help to perform
analytical calculations and Monte-Carlo simulations. They are probably
in the same universal class.

In the forthcoming publication we shall present more details concerning
the inequality (3), which plays crucial role in this theory and
establishes
an
absolute minimum of the Steiner functional $A(M)$. On the plane it
reduces
to the well known isoperimetric inequality [8]. The isoperimetric
inequality has very long history [8-10] and in the case of a
curve on the plane it states, that $L^{2} \geq 4 \pi S$, where $L$ is the
perimeter of the curve and $S$ is the enclosed area, equality takes place
only for circle.
As it already was mentioned,the
inequality (3) is also reduced to equality only for the sphere. In this
theory the sphere plays the role of instanton [1,17,18]. The
decomposition (42) has also
deep geometrical origin and is connected with linear property of $A(M)$.

Finally let us project $A(M)$ onto one- and two-dimensional spaces.
In $R^2$ the image of the squeezed surface is reduced to a polygon with
the
vertexes $X_{1},..,X_{\vert T \vert}$. The action $A(M)$ becomes
proportional
to the length of the "shadow" polygon sides, but the difference with the
gaussian model is that in this case only sides, attached to round curves,
are
distributed (the surface squeezed
without crinkles). In principle one can consider this picture as the new
version of the gaussian model in $2d$. In $R^1$ $A(M)$ is reduced to the
diameter $r$ of the vertexes.

We gratefully acknowledge conversations with R.Ambartzumian,
I.Batalin,E.Floratos, G.Flume, A.Kaidalov, B.M\"uller, H.Nielsen and  
G.Sukiasian.

One of the authors (G.S.) is thankful to R.Ambartzumian, who twenty years
ago introduced him into the beautiful Buffon's problem of the needle and
problems of geometrical probabilities, to W.Greiner for discussions and
warm hospitality at Frankfurt University and to E.Binz for the useful
comments and help during his stay at Mannheim University.
This work was done under financial support of Alexander von Humboldt
Foundation.
\medskip

\centerline{\it References} 
\vspace{1cm}

1. R.V.Ambartzumian,G.K.Savvidy,K.G.Savvidy,G.S.Sukiasian.
   Phys.Lett.B275(1992)99.

2. J.Ambjorn,D.Durhuus,J.Frohlich.Nucl.Phys.B257 (FS14) (1985) 433.

3. K.Wilson.Phis.Rev.D10(1974)3445.

4. A.Billoire,D.J.Gross,E.Marinari. Phys.Lett.B139(1984)75.

5. V.A.Kazakov. Phys.Lett.B150(1985)282.

6. F.David.Nucl.Phys.B257 (FS14) (1985)45.

7. J.Ambjorn,D.Durhuus.Phys.Lett.B188(1987)253.

8. Zenodor.${\Pi\epsilon\rho\iota}$ ${\iota\sigma o\pi\epsilon\rho\iota
   \mu\epsilon\tau\rho\omega\nu}$ ${\sigma\chi\eta\mu\alpha\tau\omega
   \nu}.$ 150 B.C.

9. J.Steiner. \"Uber Maximum und Minimum bei den Figuren in der Ebene,
     auf der Kugelfl\"ache  {\obeylines{~~~
     und im Raume \"uberhaupt. ~Gesammelte Werke. Band.2(Berlin,1882), S.177.}

10. H.Minkowski. Volumen und Oberfl\"ache.Math.Ann.B57(1903)447.

11. J.Steiner. \"Uber parallele Fl\"achen.Gesammelte Werke. Band.2 (Berlin,1882), S.171-176.

12. W.Blaschke. Kreis und Kugel,(Berlin,1956).

13. A.Cauchy. \'Memoire sur la rectification des courbes et la quadrature des surfaces courbes. 
                 Mem. Acad. Sci. Paris  22 (1850) 3.

14.L.A.Santalo.Integral geometry and geometric probability, (Reading,Mass.Addison.Wesley,1976).

15. R.V.Ambartzumian. Combinatorial integral geometry,  (Chichester.J.Wiley and Sons,1982).

16. S.S.Chern. Differential geometry and integral geometry.
   in:Proc.Internat.Congr.Math., Edinburgh, 1958 (Cambridge Univ.Press.,1960) p.441-449.

17. B.Biran,E.G.Floratos,G.K.Savvidy.Phys.Lett.B198(1988)328.

18. G.K.Savvidy. Symplectic and large-N gauge theories.In:
   Vacuum structure in intense fields, eds. H.M.Fried and B.Muller (ASI Series B255,1991) p.415.

19. J.Ambjorn, B.Durhuus, J.Fr\"ohlich and P.Orland.Nucl.Phys.B270(1986)457.

20. J.Ambjorn, B.Durhuus, J.Fr\"ohlich and T.Jonsson.Nucl.Phys.B290 (1987)480. 
   J.Ambjorn, B.Durhuus and T.Jonsson. Nucl.Phys.B316(1989)516. 
   J.Ambjorn, A.Irb\"ack, J.Jurkiewiecz  and B.Petersson. The theory of dynamical random surfaces
   with extrinsic curvature. NBI-HE-92-40.

\end{document}